\documentclass[twocolumn,twoside,slac_two]{revtex4}
\usepackage{graphicx}
\usepackage{fancyhdr}
\usepackage{pstricks}
\newcommand{\bb}{\bar{b}}
\newcommand{\bbb}{b\bar{b}}
\newcommand{\qqb}{q\bar{q}}
\newcommand{\qb}{\bar{q}}
\newcommand{\ra}{\rightarrow}
\newcommand{\ind}{{\rm d}}
\def\nn{\nonumber}
\def\Ref#1{Ref.~\cite{#1}}
\def\Refs#1{Refs.~\cite{#1}}

\def\pt{{p_{\rm T}}}
\def\ptcut{p_{\rm T}^{\rm cut}}
\begin{document}

\title{Electroweak corrections to b-jet and di-jet production}


\author{A. Scharf}
\affiliation{Department of Physics, State University New York at Buffalo, Buffalo, NY 14260, USA}

\begin{abstract}
Simultaneously with the turning-on of the Large Hadron Collider (LHC) the data taking for jet production rates will start. 
The study of these production rates is an important test of the Standard Model in the new energy regime accessible at the LHC. 
In order to discriminate possible extensions of the Standard Model accurate theoretical predictions are needed. 
We investigate the next-to-leading order weak corrections for bottom-quark jet and di-jet production at the LHC and find detectable effects 
in transverse momentum distributions. 
\end{abstract}

\maketitle

\thispagestyle{fancy}


%%%%%%%%%%%%%%%%%%%%%%%%%%%%%%%%%%
\section{Introduction}
With the start of the Large Hadron Collider (LHC) a new energy regime is accessible, either to confirm
the Standard Model (SM) or to verify new physics at the TeV scale. Famous possible 
SM extensions are heavy gauge bosons (e.g. $Z'$), Supersymmetry or 
Kaluza-Klein resonances in models with extra dimensions. Beside the outstanding discovery of the Higgs boson, processes involving top-quarks, 
gauge bosons of the weak interaction and jets are of particular interest.
The experimental identification will rely on their characteristic decay products with leptons or (bottom-) jets as characteristic examples.
In addition to this SM processes, (bottom-) jets are also involved in many decays originating from signals for physics beyond the SM. 
In particular new resonances, e.g. heavy gauge bosons, decaying into $b$-jets or light quark jets require a detailed SM-based prediction to prove a possible 
deviation from the SM.
This needs a detailed theoretical understanding of the corresponding 
processes like bottom-quark pair, single bottom-quark and di-jet production. 
These processes were studied in the past. 
The differential cross section for bottom-quark pair production is known to
next-to-leading order (NLO) accuracy in quantum chromodynamics (QCD)
\cite{Nason:1989zy,Beenakker:1988bq}. For massless single bottom production and di-jet production the NLO QCD
can be found in \Refs{Ellis:1985er,Giele:1993dj}.\\
It is well known that weak corrections can also be significant because
of the presence of possible large Sudakov logarithms. These effects were studied intensively
for several processes like weak boson and top-quark pair production
\cite{Kuhn:2005az,Kuhn:2005gv,Kuhn:2007cv,Moretti:2006nf,Moretti:2006ea,Beenakker:1993yr,Kuhn:2005it,Kuhn:2006vh,Bernreuther:2006vg}.  
Earlier work on Sudakov logarithms in four-fermion processes can be found in
\Refs{Kuhn:1999de,Kuhn:1999nn,Kuhn:2001hz,Feucht:2004rp,Jantzen:2005az,Bec,CiaCom,Fad,Bec2,DenPoz}.
A study for $b$-jet production can be found in \Ref{Kuhn:2009nf}. 

%%%%%%%%%%%%%%%%%%%%%%%%%%%%%%%%%%
\section{Leading order processes}
\label{sec:leadingorder}
\begin{figure}[!htb]
  \begin{center}
    \leavevmode
    \includegraphics[width=50mm]{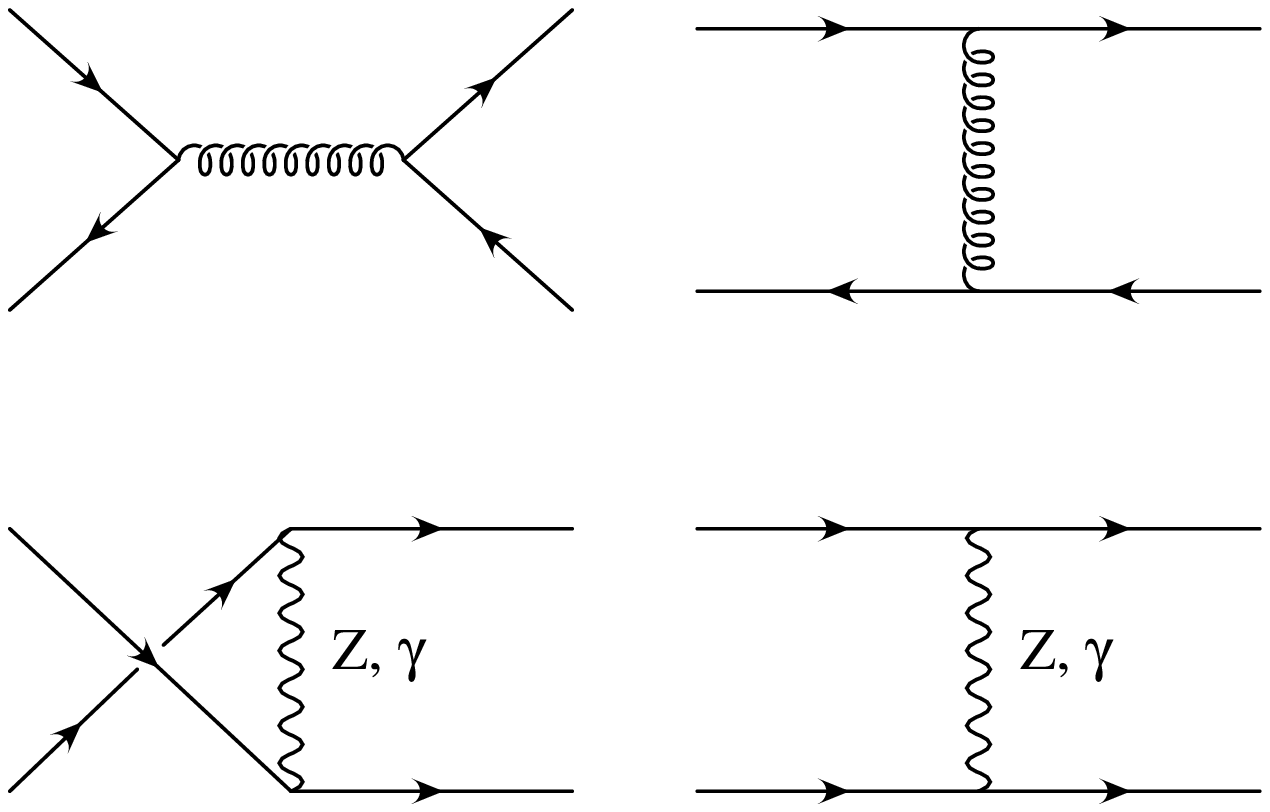}
    \caption{Sample Born diagrams for quark-induced processes.}
    \label{fig:quarkborns}
    \vspace*{2cm}
    \includegraphics[width=80mm]{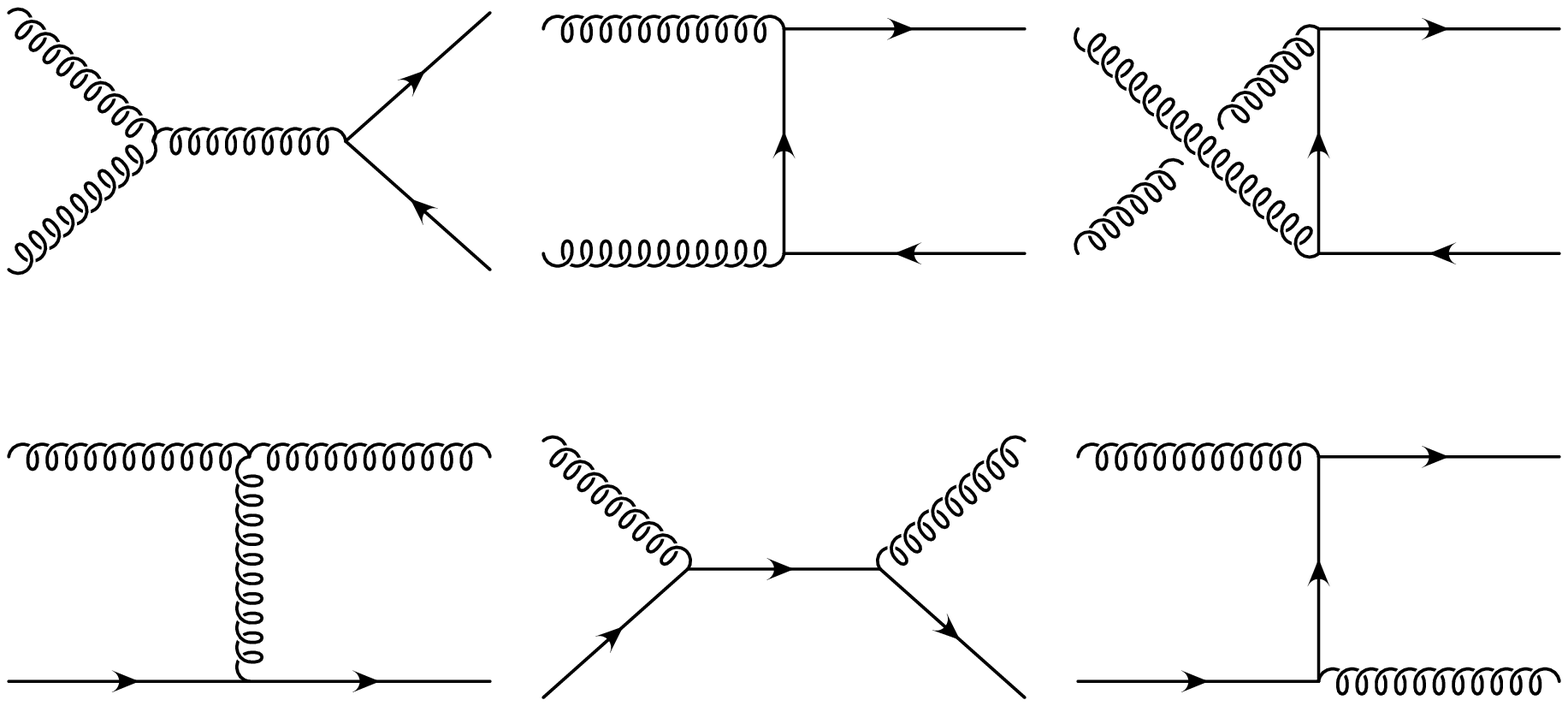}
    \caption{Sample Born diagrams for gluon-induced processes.}
    \label{fig:gluonborns}
  \end{center}
\end{figure}

In the following processes with initial state photons are neglected and
we distinguish between quark-induced processes (with two quarks in the initial state, one of these being $u,d,c,s$), 
gluon-induced contributions (with one or two gluons in
the initial state) and in the case of $b$-jet production also pure bottom-quark induced processes. 
The LO gluon induced processes proceed through QCD amplitudes only while the four quark processes 
can be seperated into QCD $O(\alpha_S^2)$, electroweak QCD $O(\alpha^2)$ and mixed QCD-electroweak $O(\alpha\alpha_S)$ 
contributions. Sample diagrams
are shown in Fig.~\ref{fig:quarkborns} and Fig.~\ref{fig:gluonborns}.  
Here the light quarks (denoted generically by $q$ and $\qb$) and the bottom-quark
($b$ and $\bb$) are taken as massless. Since jets close to the beam pipe escape detection, we require a minimal 
transverse momentum $\ptcut$ of 50 GeV. The leading order cross section is then obtained from 
\begin{eqnarray}
\ind\sigma_{H_1,H_2\ra X} &=& \sum_{i,j}\int_0^1 \ind x_1\int_0^1 \ind x_2
%\Theta(x_1x_2-\tau_{i,j})
f_{i/H_1}(x_1)f_{j/H_2}(x_2) \nn\\
&\times& 
\ind\hat{\sigma}_{i,j\ra X}(x_1P_1,x_2P_2)\:\:\:\Theta(\pt > \ptcut),\nn\\
\label{eq:Hadron-WQ}
\end{eqnarray}
where the factorization scale dependence is suppressed and $x_1$ and $x_2$ are the
partonic momentum fractions. The parton
distribution functions (PDF's) for parton $i$ in hadron $H$ are denoted by $f_{i/H}$. 
The sum runs over all possible parton configurations $(i,j)$ in the initial state. 
\subsection{$b$-jet production}
\begin{figure*}[!htb]
  \begin{center}
    \leavevmode
    \hspace*{-0.4cm}
    \begin{minipage}{15cm}
    \includegraphics[width=0.49\textwidth]{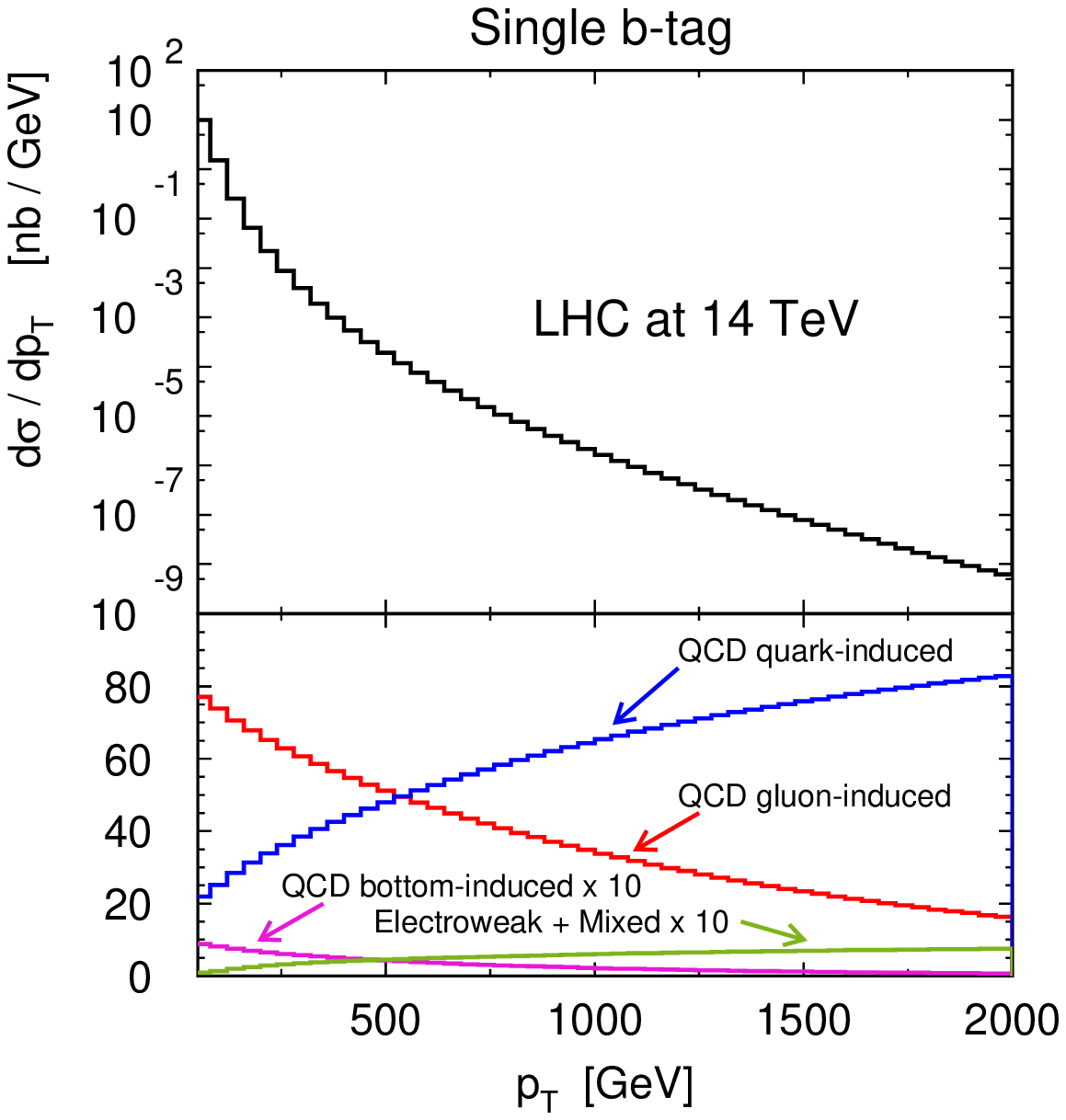}
    \includegraphics[width=0.49\textwidth]{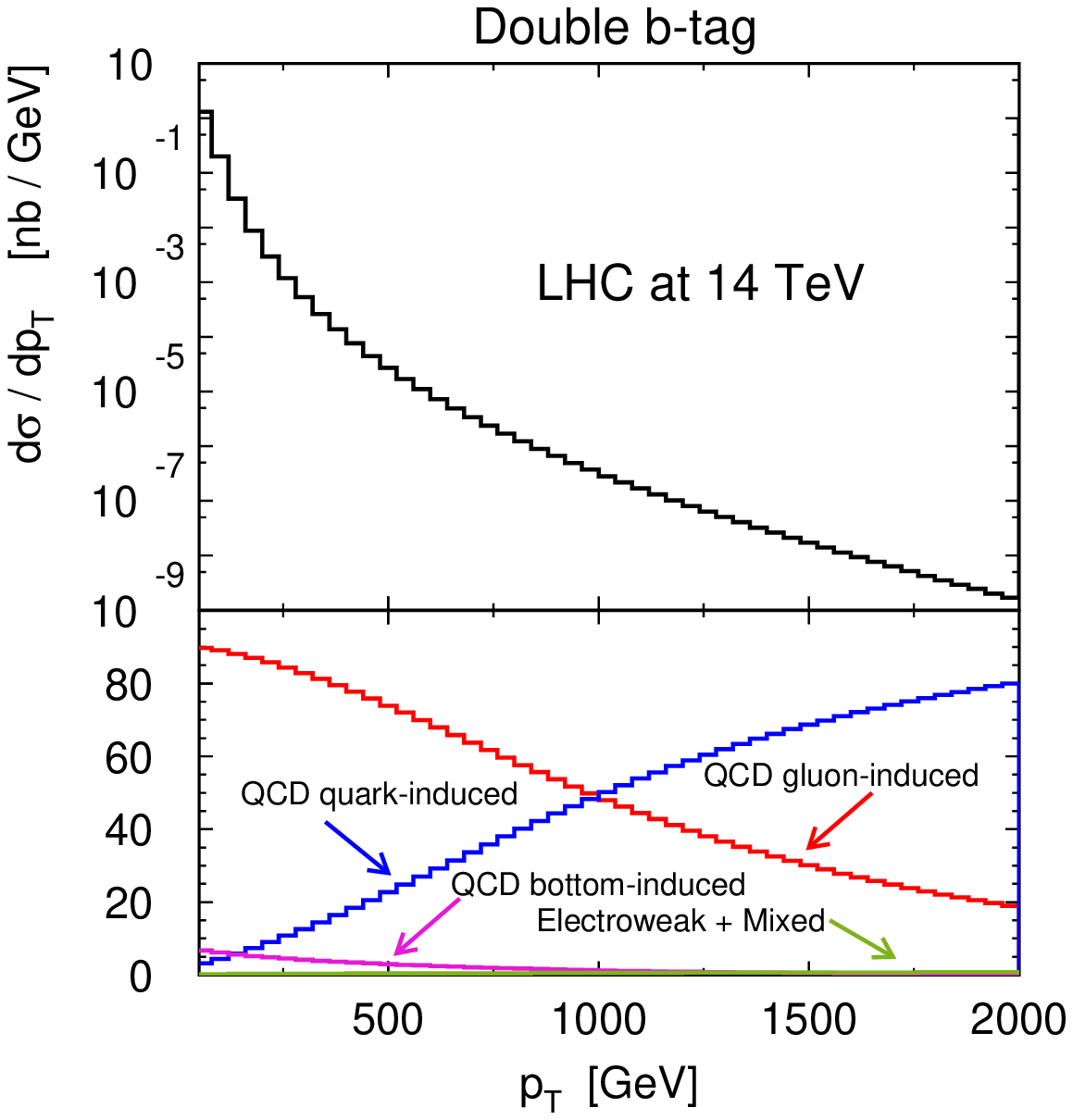}
    \end{minipage}
    \rput(-8.7,2.8){\large (a)}
    \rput(-1.25,2.8){\large (b)}
    \caption{Differential cross section as a function of
       $\pt$ for single $b$-tag events (upper left figure), double $b$-tag events (upper right figure) 
       at the LHC ($\sqrt{s} = 14 $ TeV) and the relative composition normalized to the Born cross section (lower
       figures).}
    \label{fig:pt-bjet-lhc}
  \end{center}
\end{figure*}

For $b$-jet production three types of partonic processes will be distinguished at leading order:
\vspace*{-0.2cm}
\begin{eqnarray}
{\rm quark-induced}:&&\qqb\ra\bbb,\:\:\bb q\ra\bb q,\:\: b\qb\ra b\qb,\nn\\
&&bq\ra bq,\:\: \bb\qb\ra\bb\qb, \nn\\
{\rm gluon-induced}:&&  gg\ra\bbb,\:\: bg\ra bg,\:\: \bb g\ra\bb g, \nn\\
{\rm bottom-induced}:&& \bbb\ra\bbb,\:\: bb\ra bb,\:\: \bb\bb \ra\bb \bb.\nn
\label{eq:lo-list}
\end{eqnarray} 
Furthermore we distinguish between single $b$-tag and double $b$-tag events.
The LO $\pt$-distribution for single $b$-tag events at the LHC is presented in Fig.~\ref{fig:pt-bjet-lhc}(a).
The gluon-induced processes dominate in the low energy regime 
($\pt < 500$ GeV). For $\pt$ larger than 500 GeV the 
quark-induced processes take over and finally dominate the
distribution. This "cross over" of gluon- and quark-induced contributions is a consequence of 
the different behaviour of LHC quark and gluon luminosities.  
The relative contributions from the exchange of electroweak bosons and 
bottom-induced processes are always below $1\%$ and therefore negligible.
A similiar picture is observed for the differential cross section for
double $b$-tag events at the LHC (Fig.~\ref{fig:pt-bjet-lhc}(b)). Here the "cross over" of quark- and 
gluon-induced contributions is around $\pt = 1$ TeV. 
For low $\pt$ the pure bottom-induced processes are as important as quark-induced contributions. 
This seems surprising, because the parton luminosities of bottom-quarks
in a proton should be highly suppressed compared
to the light flavours. There are two reasons for the
relatively large cross section of the purely bottom-induced processes. 
First, the partonic cross sections of $bb$ and 
$\bbb$ scattering are strongly enhanced for large $z$ values. In this region the parton 
processes with bottom-quarks in the initial state 
can be several orders of magnitude larger than the quark--antiquark-induced process. Second, the
bottom-quark PDF is essentially obtained by multiplying the gluon distribution in the
proton with the splitting function of a gluon into a bottom-quark pair. 
Because of the high gluon luminosities at low energies, 
the bottom-quark PDF becomes of the order of a few percent relative to the PDF's of the light
flavours. This, together with the large partonic contributions is responsible for
the relatively large bottom-induced differential cross section. The argumentation implies 
that the main contribution from $bb$, $\bb\bb$ and $\bbb$ scattering comes from the low $\pt$ region,
while for high $\pt$ values these effects are small. It might be interesting to study whether the
$b$-PDF could be further constrained using $\bbb$ production at low $\pt$. 
\\
As shown above, the leading order contributions from electroweak
gauge boson exchange are always negligible for the study of $b$-jet production. This is also true in the context of
NLO corrections with an expected size of serveral percent relative to
the leading order distributions. Moreover, we have shown that the QCD contributions from processes with two bottom-quarks
in the initial state are unimportant for the study of $\pt$-distributions at large transverse momentum. 
In particular with regard to the Sudakov logarithms becoming important at high energies this approximation is 
justified. Consequently the weak $O(\alpha)$ corrections to $bb\ra bb$, 
$\bb\bb\ra\bb\bb$ and $\bbb\ra\bbb$ will not be taken into account. For further details we refer 
to \Ref{Kuhn:2009nf}. 
\subsection{Di-jet production}
\begin{figure}[!htb]
  \begin{center}
    \leavevmode
    \includegraphics[width=80mm]{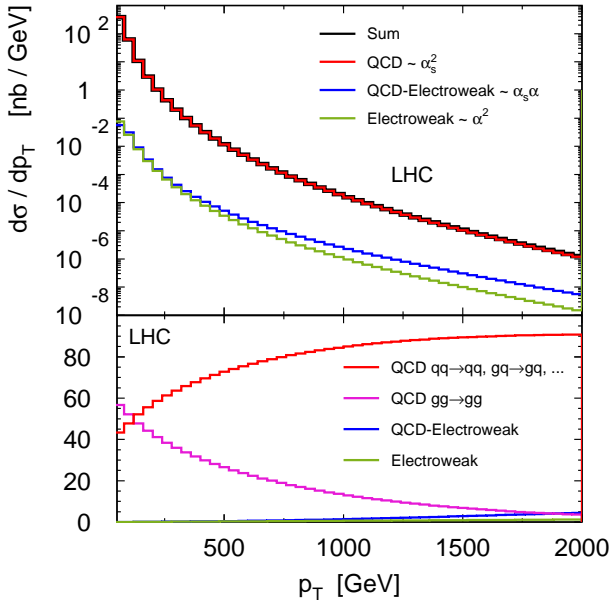}
    \caption{Differential cross section as a function of $\pt$ for di-jet events(upper figure) at the LHC ($\sqrt{s} = 14 $ TeV)  
      and the relative compsosition normalized to the Born cross section (lower figure).}
    \label{fig:di-jetLO}
  \end{center}
\end{figure}

For di-jet production the partonic final state consists of light quarks and/or gluons. 
Similiar to $b$-jet production we distinguish between quark- and gluon induced processes.
 \begin{eqnarray}
{\rm quark-induced}:&&\qqb\ra q'\qb',\:\:q\qb'\ra q\qb',\:\:qq'\ra qq',\nn\\
&&\qb\qb'\ra \qb\qb',\:\:\qqb\ra \qqb,\:\:qq\ra qq,\nn\\
&&\qb\qb\ra \qb\qb,\:\:\qqb\ra gg\nn\\
{\rm gluon-induced}:&&  gg\ra\qqb,\:\: qg\ra qg,\:\: \qb g\ra\qb g,\nn\\
&& gg\ra gg.\nn
\label{eq:lo-list2}
\end{eqnarray} 
The LO $\pt$-distribution and its composition at the LHC is shown in Fig.~\ref{fig:di-jetLO}.
The differential cross section is dominated by contributions coming from the
QCD processes. The contributions of $O(\alpha\alpha_S)$ and $O(\alpha^2)$ are at the
permille level for transverse momenta up to 1 TeV. For higher $\pt$ values the electroweak 
contributions remain insignificant while the mixed QCD-electroweak contributions yield up to 
5\% to the differential cross section at $\pt = $ 2 TeV. In context of the Sudakov Logarithms the four gluon 
process will not receive weak corrections within the SM at NLO accuracy. Therefore in the lower figure of Fig.~\ref{fig:di-jetLO} the
relative impact of this process on the $\pt$-distribution is shown in magenta. 
For $\pt < 100$ GeV $gg\ra gg$ provides the major contribution to the differential cross section. With increasing
$\pt$ values the relative contribution drops but remains non-negligible for the depicted $\pt$ region. For $\pt > 1.5$ TeV 90\%
of the cross section is delivered by the remaining QCD processes, which receive NLO weak corrections. \\
The discussion shows that the contributions of $O(\alpha^2)$ are always negligible for di-jet events at the LHC.
Investigating the NLO $O(\alpha\alpha_S^2)$ corrections to di-jet production the $O(\alpha\alpha_S)$ 
and $O(\alpha_S^2)$ contributions can not be seperated and therefore QCD and weak corrections must be calculated. Moreover it 
was shown, that the relative contribution of the partonic channel $gg\ra gg$ is of the order of 5-20\% in the high energy regime, which justifies 
the calculation of the weak corrections. 
\section{Weak corrections of $O(\alpha\alpha_S^2)$}
\label{sec:NLO}
\begin{figure}[!htb]
  \begin{center}
    \includegraphics[width=80mm]{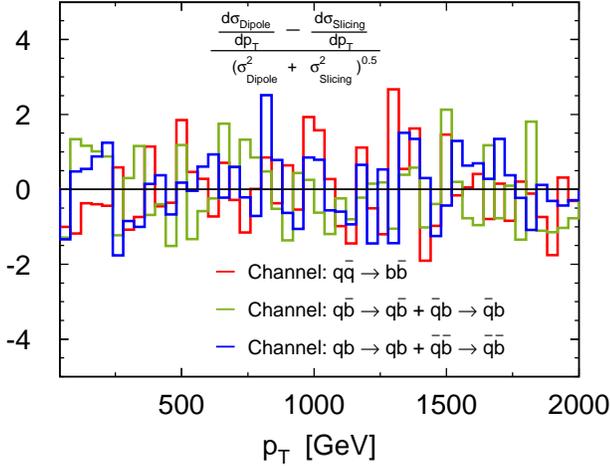}
     \caption{Difference between results based on dipole formalism and phase space slicing for the differential cross sections 
       for the different quark-induced channels in terms of standard deviations.
     For the comparison only the relevant IR contributions (boxes and real corrections) were taken into account.}
     \label{fig:cut-study}
  \end{center}
\end{figure}

\begin{figure}[!htbp]
  \begin{center}
    \leavevmode
    \vspace*{0.5cm}
    \includegraphics[width=8cm]{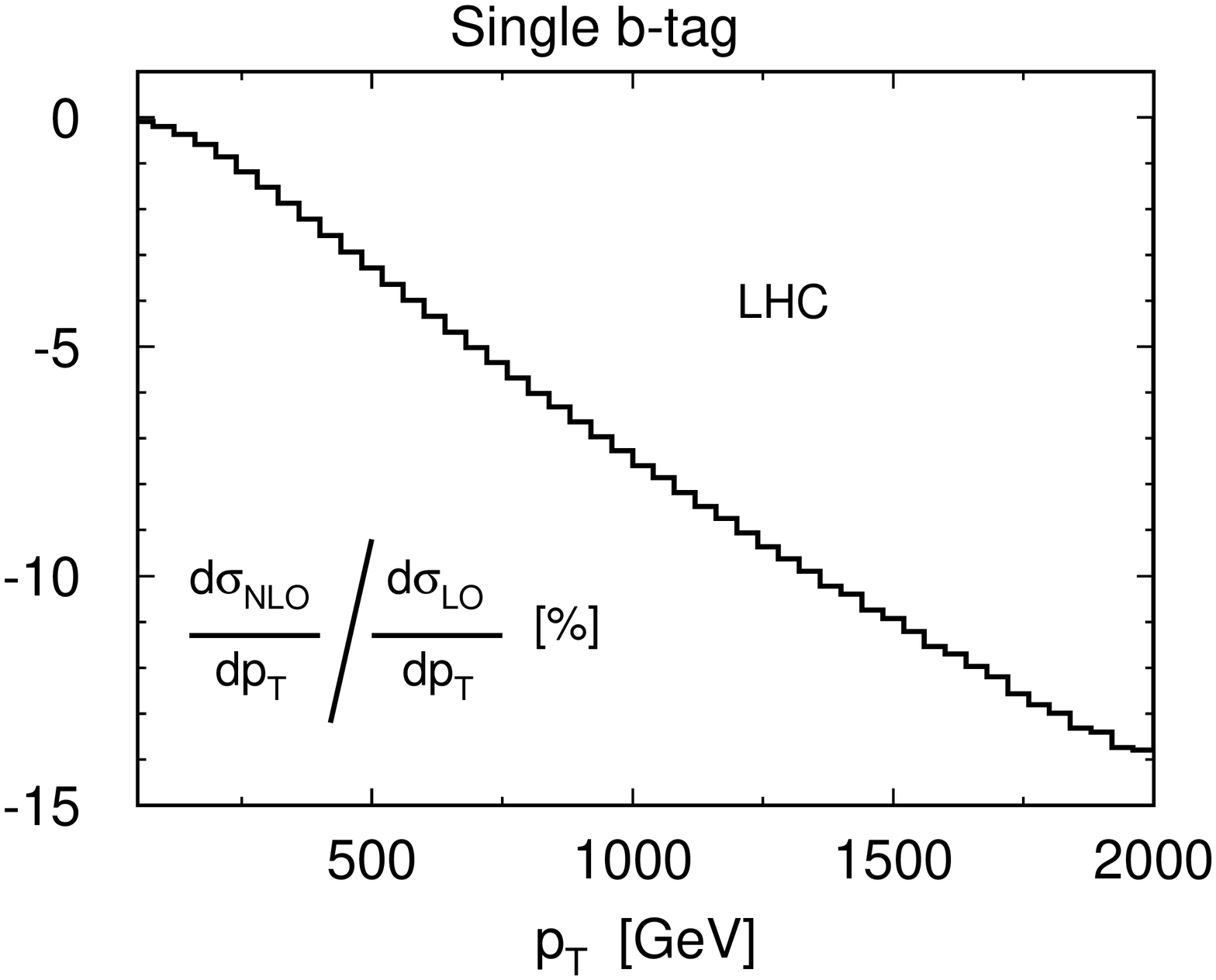}
    \includegraphics[width=8cm]{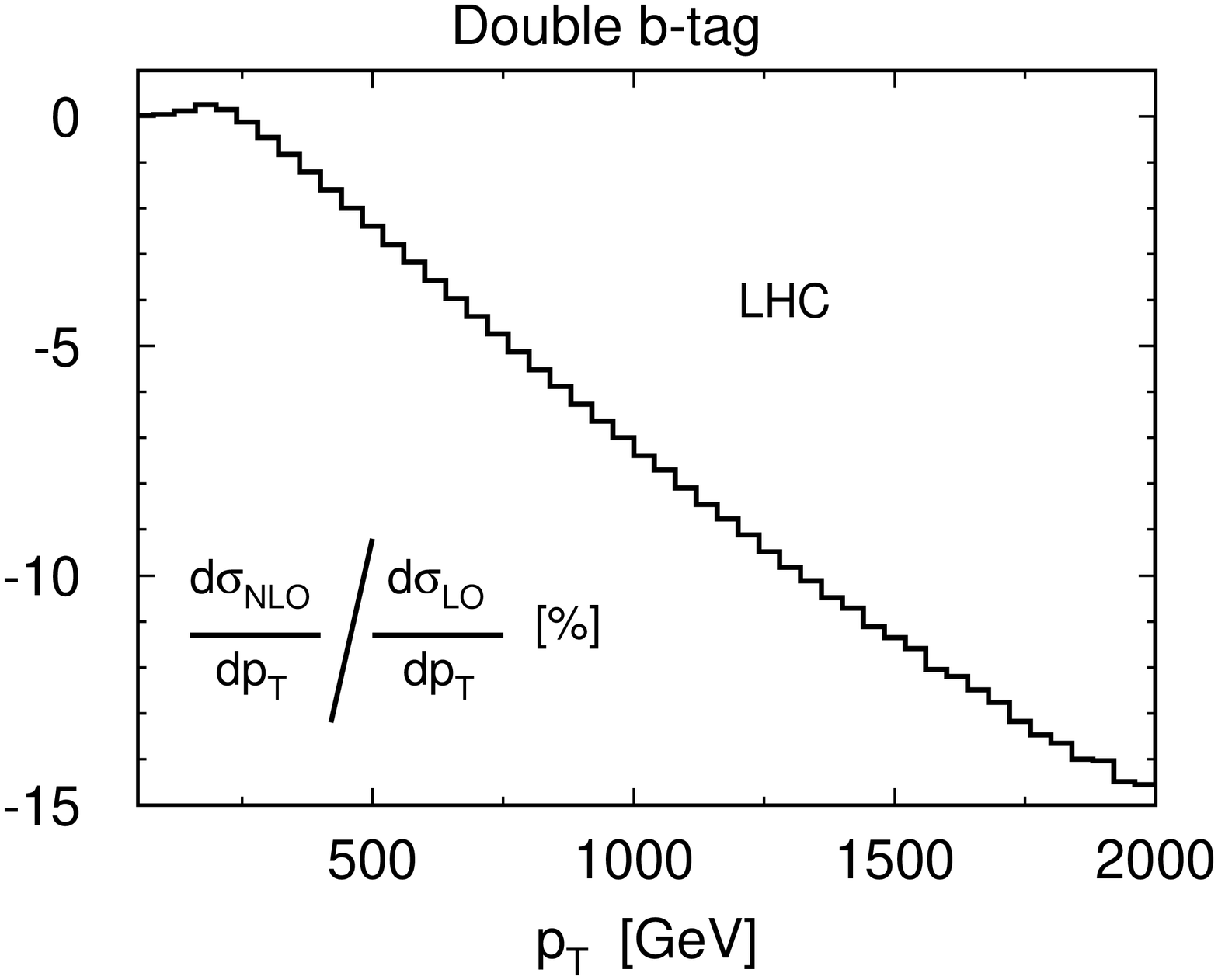}
     \caption{Relative weak corrections for single $b$-tag (upper figure) and 
       double $b$-tag (lower figure) events at the LHC ($\sqrt{s} = 14$ TeV).}
     \label{fig:pt-nlo-bjet-relative-LHC}
  \end{center}
\end{figure}

\begin{figure}[!htbp]
  \begin{center}
    \leavevmode
    \vspace*{0.5cm}
    \includegraphics[width=8cm]{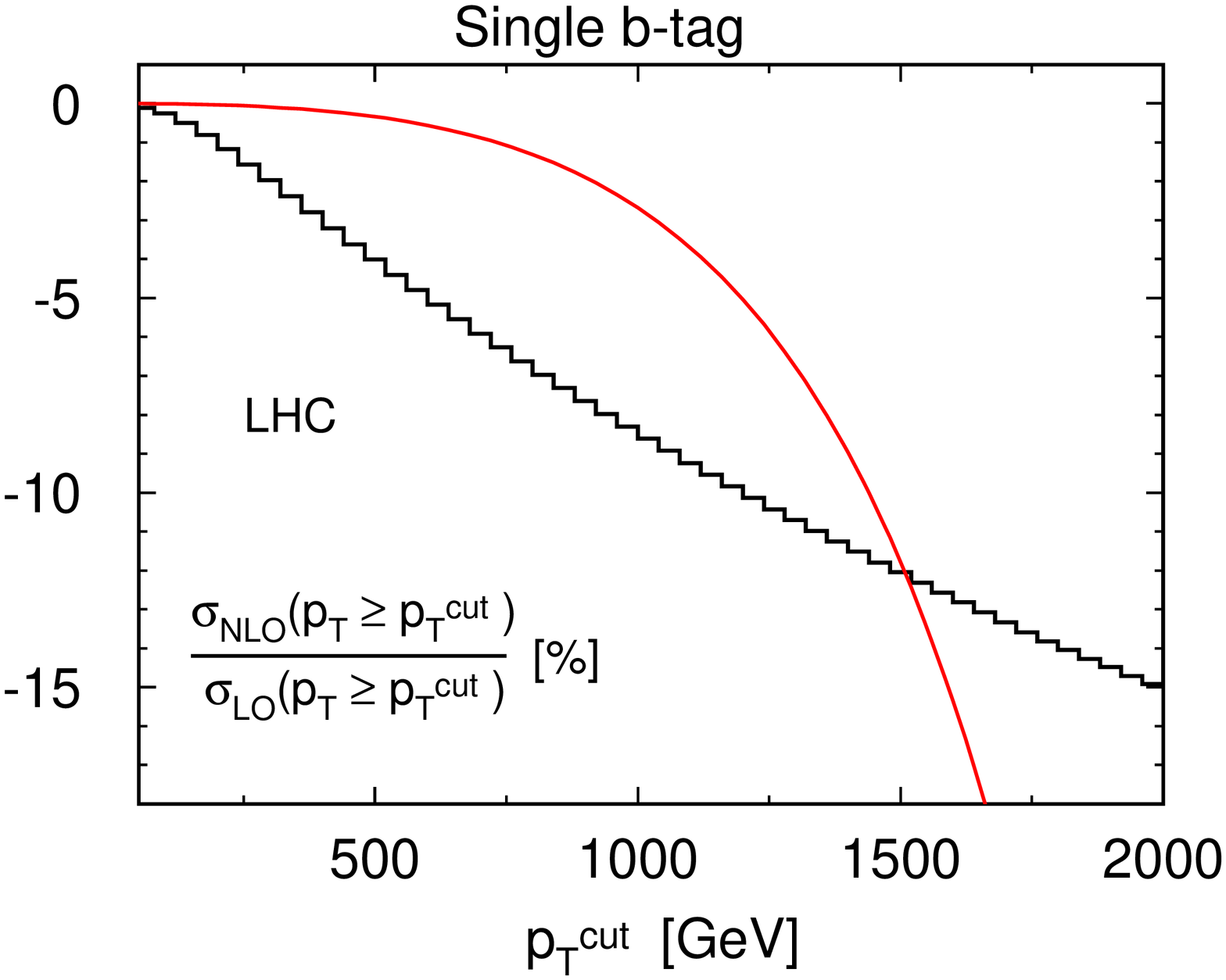}
    \includegraphics[width=8cm]{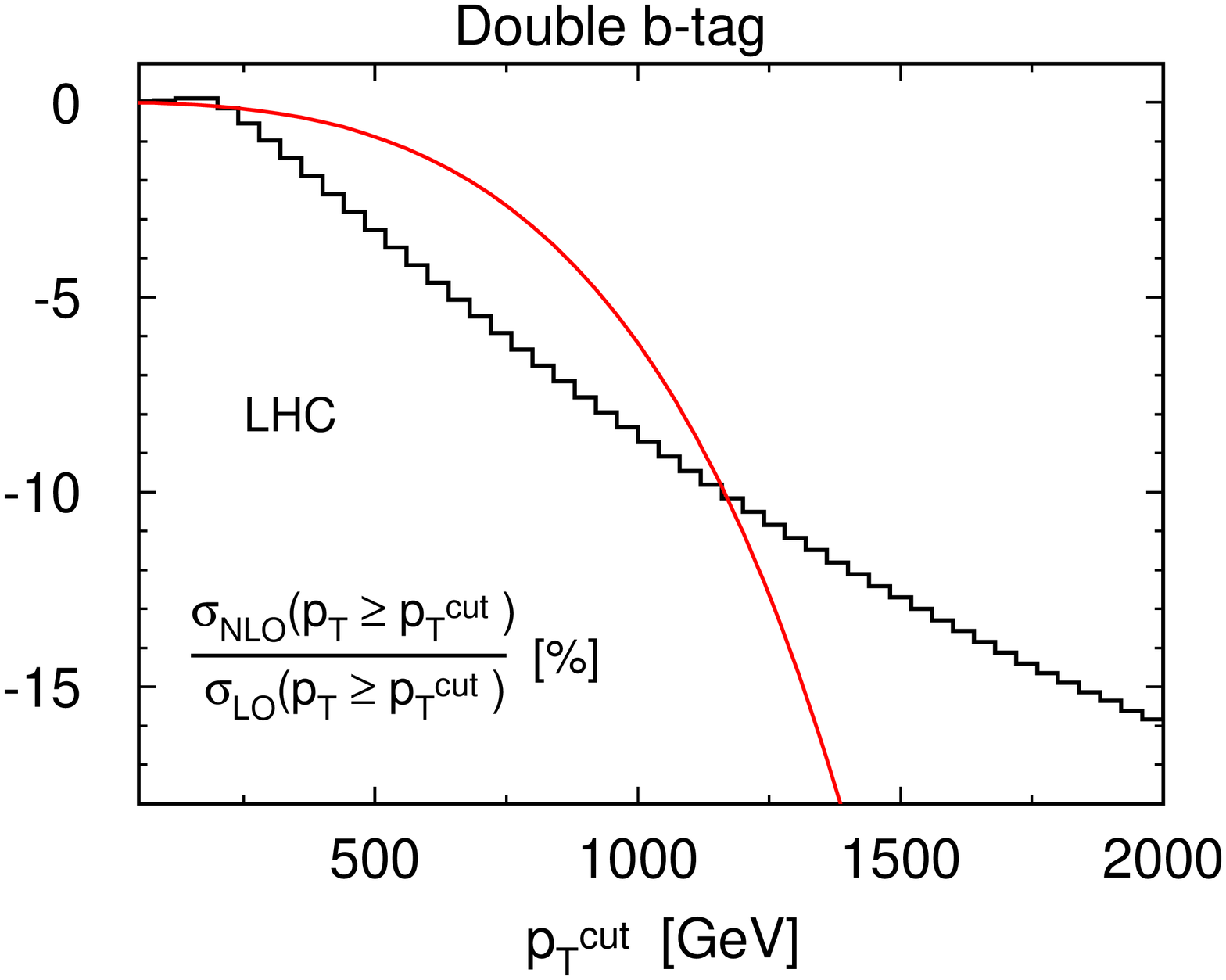}
     \caption{Relative corrections to the cross section for $\pt > \ptcut$ at the LHC for single $b$-tag (upper figure) and 
double $b$-tag events (lower figure). The red lines give an estimate on the statistical uncertainty as described in the text.}
     \label{fig:pt-cut-nlo-bjet-relative-LHC}
  \end{center}
\end{figure}

\begin{figure*}[!htbp]
  \begin{center}
    \leavevmode
    \includegraphics[width=0.49\textwidth]{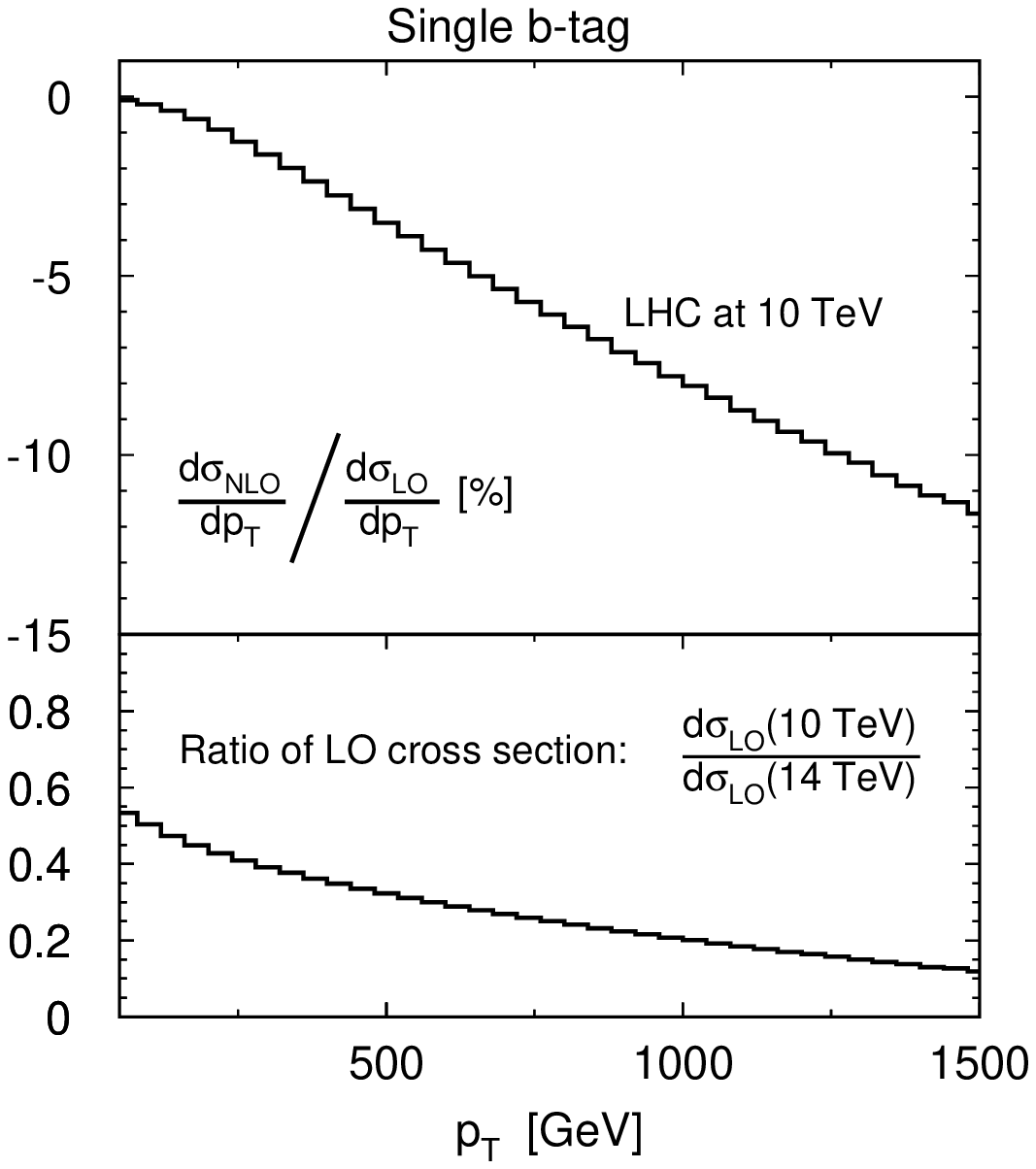}
    \includegraphics[width=0.49\textwidth]{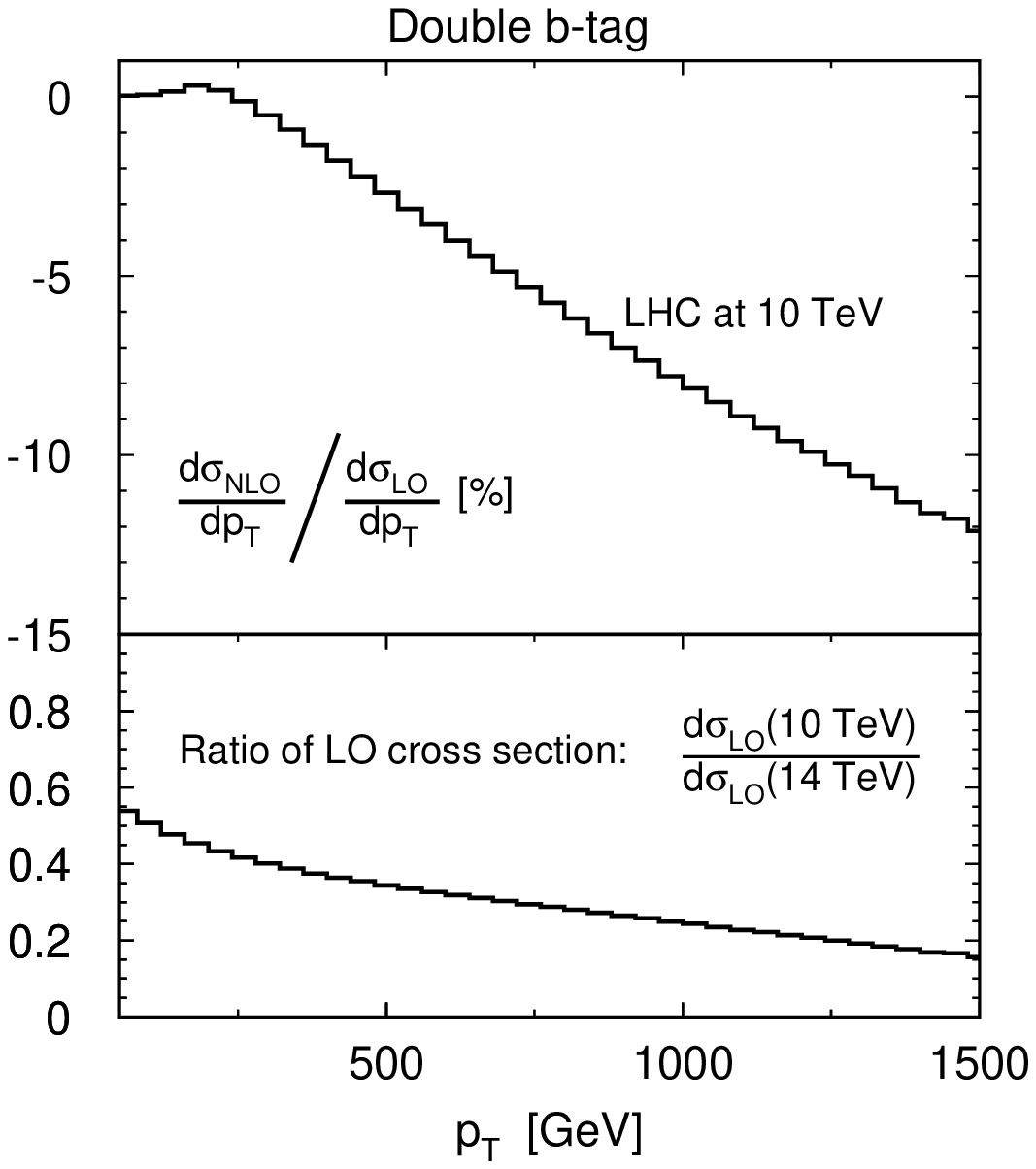}
     \caption{Relative corrections for single $b$-tag (a) and double $b$-tag events (b) at the LHC operating with a center-of-mass energy $\sqrt{s} = 10$ TeV (upper figures) 
       and the ratio of the differential leading order cross sections at the LHC operating at 10 TeV and 14 TeV 
       $\left(\ind\sigma_{\rm LO}(10\:{\rm TeV})/ \ind\pt\right)/\left(\ind\sigma_{\rm LO}(14\:{\rm TeV})/ \ind\pt\right)$ (lower figures).}
     \label{fig:pt-LHC10}
  \end{center}
  \rput(-1.3,10.45){\large (a)}
  \rput(7.2,10.45){\large (b)}
\end{figure*}

\begin{figure}[!htbp]
  \begin{center}
    \leavevmode
    \includegraphics[width=0.49\textwidth]{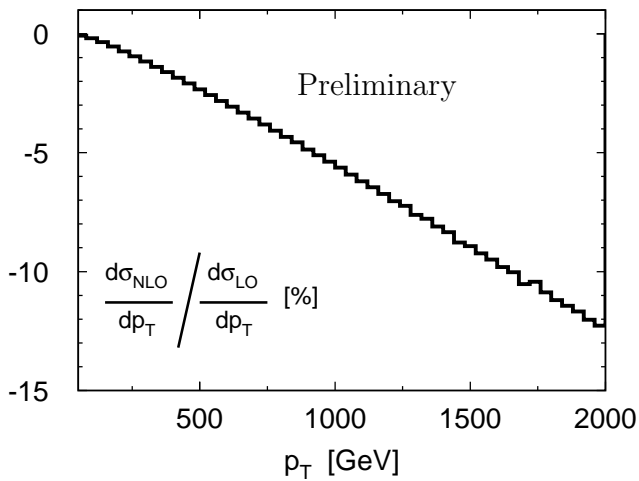}
     \caption{Relative corrections for di-jet events at the LHC ($\sqrt{s} = 14$ TeV).}
     \label{fig:pt-dijet-nlo}
  \end{center}
  \rput(0.8,7){\large Preliminary}
\end{figure}

As before we subdivide the partonic channels in contributions from quark-induced
and gluon-induced processes. For the calculation of the next-to-leading order
weak corrections the `t Hooft-Feynman gauge with gauge parameters set to
1 is used. The longitudinal degrees of freedom of the massive gauge bosons $Z$ and $W$
are thus represented by the goldstone fields $\chi$ and $\phi$. Since all incoming and outgoing
partons are massless there are no
contributions from the Goldstone boson $\chi$ and the Higgs boson (also not $\phi$ for di-jet production). 
Ghost fields do not contribute at the order under consideration. The virtual contributions to quark-induced processes contain 
infrared and ultraviolet singularities, while the gluon-induced ones are infrared finite. 
Concerning $O(\alpha)$ corrections to QCD processes only wave function renormalisation is needed and no 
mass or coupling renormalisation has to performed. For the QCD corrections to $O(\alpha\alpha_S)$ processes 
the renormalisation is performed in the $\overline{ \rm MS}$ scheme. The quark-induced contributions consist of virtual and
real corrections and to handle the IR singularities we use the dipole subtraction formalism by \Ref{Catani:1996vz}. In addition the 
results for $b$-jet production were cross-checked by implementing a phase space slicing method \cite{Beenakker:1988bq}.
The comparison between slicing and subtraction method is shown in Fig.~\ref{fig:cut-study}
\\
For the subsequent discsusion $\sqrt{s} = 14$ TeV will be adopted.
The relative corrections for $b$-jet production to the leading order $\pt$-distributions are shown in Fig.~\ref{fig:pt-nlo-bjet-relative-LHC}.
For single $b$-tag events (upper figure) the relative corrections are always negative and of the order of a few permille up to $-1\%$ for 50 GeV $< \pt <$ 250 GeV. 
For 250 GeV $< \pt< $ 1 TeV the weak NLO contributions vary between $-1\%$ and $-8\%$, compared to the leading order distribution, a consequence of the Sudakov logarithms.
In the high energy regime ($\pt> 1$ TeV) the relative corrections amount to $-10\%$ and will even reach $-14\%$ for $\pt = 2$ TeV. \\
The lower figure in Fig.~\ref{fig:pt-nlo-bjet-relative-LHC} shows the relative corrections for double $b$-tag events at the LHC. 
Despite the strong suppression of $\qqb\ra\bbb$ at leading order (see Fig.~\ref{fig:pt-bjet-lhc}(b)) a small remnant of the enhancement  
from virtual top-quarks is visible in the double $b$-tag case. 
With increasing $\pt$ the Sudakov logarithms dominate the shape of the weak NLO contributions 
and yield relative corrections between $-1\%$ and $-7\%$ (250 GeV $< \pt< $ 1 TeV). At the highest $\pt$-values considered relative corrections 
up to $-14\%$ are observed. 
\\
Fig.~\ref{fig:pt-cut-nlo-bjet-relative-LHC} (upper figure) shows the integrated cross section for single $b$-tag events at the LHC together with an estimate of the
statistical error based on an integrated luminosity of 200 ${\rm fb}^{-1}$. The same composition in shape
and magnitude is observed as for the differential distribution. The statistical error estimate matches the size of the weak corrections up to $\pt = 1.5$ TeV. 
For higher $\pt$-values the rate drops quickly and it will be difficult to observe the effect of the weak corrections. For double $b$-tag events (lower figure) 
we find again the smoothing of the $t\bar{t}$-threshold for the relative weak corrections, while the composition of the curve for $\pt> 250$ GeV is very similiar to
the already discussed differential distribution. Considering the statistical error, the slight increase from virtual top-quarks will not be observable at the LHC. 
For $\pt$-values between $250$-$1000$ GeV the weak corrections are larger than the statistical error, above $\pt =1$ TeV they are comparable 
or smaller.\\ 
In Fig.~\ref{fig:pt-LHC10} we show results for the LHC operating at $\sqrt{s} = 10$ TeV. 
The absolute cross sections are by more than a factor two lower, due to the lower parton luminosities.
However the impact of the electroweak corrections is nearly the same and qualitatively similiar results are obtained for the relative NLO corrections.
\\
For di-jet production we present a preliminary result for the realtive weak corrections to the $\pt$-distribution in Fig.~\ref{fig:pt-dijet-nlo}. 
We find similiar results like for the $b$-jet case. However the magnitude of the weak corrections is smaller. For $50<\pt<1000$ GeV the relative corrections 
are negative and of the order of a few percent. For $\pt >$ 1 TeV the NLO weak corrections become of the order of $-10\%$ up to $-12\%$ at $\pt=2$ TeV. 
\section{Conclusions}
We have presented numerical predictions for the weak NLO corrections to $b$-jet and di-jet production at the LHC. 
For $b$-jet production we find the NLO weak corrections decrease the LO $\pt$-distribution by -10\% for transverse momenta around 1 TeV
for single $b$-tag as well as double $b$-tag events. These effects are larger than the anticipated statistical uncertainty. 
The preliminary result for di-jet production indicates that the weak corrections lower the LO $\pt$-distribution by several percent
for high transverse momenta at the LHC.
 \label{sec:results}

%%%%%%%%%%%%%%%%%%%%%%%%%%%%%%%%%%

\bigskip % extra skip inserted

\end{document}